%% file: main.tex
\documentclass[sigconf]{acmart}

\usepackage{subcaption}
\usepackage{multirow}
\usepackage{listings}
\usepackage{url}
\usepackage{xspace}
\usepackage[utf8]{inputenc}

\newcommand{\name}{I\textsuperscript{3}DE\xspace}

\copyrightyear{2024}
\acmYear{2024}
\setcopyright{acmlicensed}
\acmConference[IDE '24]{2024 First IDE
Workshop}{April 20, 2024}{Lisbon, Portugal}
\acmBooktitle{2024 First IDE Workshop (IDE '24), April 20, 2024, Lisbon,
Portugal}
\acmDOI{10.1145/3643796.3648461}
\acmISBN{979-8-4007-0580-9/24/04}

\begin{document}

\title{\name: An IDE for Inspecting Inconsistencies in PL/SQL Code}

\author{Jiangshan Liu}
\affiliation{
  \institution{College of Intelligence and Computing\\Tianjin University}
  \city{Tianjin}
  \country{China}}
\email{jiangshan.liu@tju.edu.cn}
\orcid{0009-0004-1215-1065}
\author{Shuang Liu}
\authornote{Shuang Liu is the corresponding author.}
\affiliation{
  \institution{School of Information\\Renmin University of China}
  \city{Beijing}
  \country{China}}
\email{Shuang.Liu@ruc.edu.cn}
\orcid{0000-0001-8766-7235}
\author{Junjie Chen}
\affiliation{
  \institution{College of Intelligence and Computing\\Tianjin University}
  \city{Tianjin}
  \country{China}}
\email{junjiechen@tju.edu.cn}
\orcid{0000-0003-3056-9962}


\input{sections/0_Abstract}





\maketitle

\input{sections/1_Introduction}
\input{sections/2_PLSQL_IC3}

\input{sections/3_IDE_Integration}
\input{sections/4_Evaluation}

\input{sections/5_Related_Work}
\input{sections/6_Conclusion}

\begin{acks}
  This work has been supported by the National Natural Science Foundation of China 61802275.
\end{acks}

\bibliographystyle{ACM-Reference-Format}



\end{document}

%% file: sections/0_Abstract.tex
\begin{abstract}
In this paper, we introduce \textbf{\name (Inconsistency Inspecting IDE)} — an IDE plugin to inspect inconsistencies in PL/SQL code. 
We first observed the potential issues, e.g., misuses or bugs,  that are introduced by the inconsistent understanding of PL/SQL semantics by PL/SQL programmers and DBMS developers,  
and propose a metamorphic testing-based  approach for inspecting such inconsistencies in PL/SQL code.
We design and implement our approach in \name, a widely usable plugin for the IntelliJ Platform. We conducted a comparative user study involving 16 participants, and the findings indicate that \name is consistently effective and efficient in helping programmers identify and avoid inconsistencies across different programming difficulties.
\end{abstract} 

%% file: sections/1_Introduction.tex
\section{Introduction}

PL/SQL serves as a procedural extension to SQL within a Database Management System (DBMS)~\cite{PostgreSQLOverview}. 
Unlike SQL, there is no standardized specification for PL/SQL~\cite{PostgreSQLPortingFromOracle}, and the  lengthy and incomplete development documentation of DBMS  creates barriers to programmers' understanding of PL/SQL.  
This could naturally lead to inconsistencies between their understanding of the PL/SQL semantics and the semantics actually implemented in the DBMS. 
In addition, our investigation~\cite{TechReport} 
conducted with a sample of 57 PL/SQL programmers and developers involved in the design and implementation of PL/SQL engines, substantiates this observation. 
The results showed  that 76.92\% of the PL/SQL programmers understood PL/SQL semantics based on their previous experience with SQL, 69.23\% of the developers implemented the PL/SQL engines (e.g., PosgreSQL~\cite{PostgreSQLOverview} and openGauss~\cite{openGaussOverview}) relying on their experience with other procedural languages and 46.15\% of the developers also rely on their own personal experience and understanding. 
Therefore, DBMS developers and PL/SQL programmers  may have inconsistent understanding about the PL/SQL semantics. 

Inconsistent understanding of PL/SQL semantics between PL/SQL programmers and DBMS developers (we use inconsistency for short in the rest of the paper) could easily result in misuse or even bugs with serious consequences. 
For instance, Figure \ref{fig:SQL-injection} illustrates that PL/pgSQL\footnote{PL/SQL originally coined by Oracle has been extended to similar languages by a range of databases. The example is adopted from PostgreSQL, which implements the PL/pgSQL language.}
code containing inconsistencies could be risky to SQL injection. In this context, PL/pgSQL
receives a {\ttfamily CHAR} type parameter (line 1) as input.
In SQL, the default length of {\ttfamily CHAR(n)} is 1, and any type conversion to an unspecified length {\ttfamily CHAR} is  truncated to the first character~\cite{PostgreSQLCHAR}. 
However, 
PL/pgSQL engine does not conduct automatic truncation on {\ttfamily CHAR} type parameters, and passes the input text as it is. 
When we pass the parameter {\ttfamily `2 OR TRUE'} (line 8) to the function, a programmer with SQL experience may expect an automatic truncation to a {\ttfamily CHAR} type, yet in PL/pgSQL, there is no such automatic truncation and thus the full-length parameter is passed and concatenated to the command in line 3 (|| is the string concatenation operator). Therefore, the \textit{where} expression is evaluated to truth, which triggers the update operation. The update set the userpass to default, which is a typical SQL injection.  

\input{sections/1_Introduction_Figure_SQL_Injection}

The identification and mitigation of potentially harmful inconsistencies lie within the purview of the programmer, presenting a challenging task~\cite{nuseibeh2001making}. One of the most viable and practical solution is to leverage \textbf{Integrated Development Environments (IDEs)} to alert programmers to these inconsistencies during their programming phase. Currently, various IDEs extensively support database development through native features or plugins~\cite{JetBrainsTools}, but no IDE has yet been able to warn programmers about such inconsistencies in PL/SQL.

\input{sections/2_PLSQL_IC3_Figure_Overview}

In this paper, we present \name, a user-friendly IntelliJ plugin that inspects inconsistencies in PL/SQL code through two modes, i.e., dynamic mode and static mode. 
The dynamic mode automatically executes the PL/SQL programs with our metamorphic testing based inspection engine and reports potential inconsistencies. 
The static mode relies on three types of pre-defined patterns to inspect the inconsistencies. 
The dynamic mode could help us inspect more inconsistencies while the static mode is effective in inconsistency inspection. 
Note that the inconsistencies inspected in the dynamic mode can be encoded in the patterns used in static mode, which enriches the patterns and at the same time maintains effectiveness. 

We conducted a user study on \name to evaluate its effectiveness in helping programmers identify and avoid inconsistencies. Participants were divided into experiment group and control group, where the experiment group completes  the task with the aid of \name, while the control group completes  the task on their own.  
Results indicate that, compared to the control group, the experiment group utilizing \name demonstrated an 87.5\% increase in correctness rate for identifying and correcting inconsistencies, with a corresponding speed improvement of 98.96\%. 

To summarize, we made the following contributions. 
\begin{itemize}
    \item We first observed the potential issues that are introduced by the inconsistent understanding of PL/SQL semantics by PL/SQL programmers and DBMS developers,   
    and propose a metamorphic testing-based   approach for inspecting inconsistencies in PL/SQL code.
    \item We developed \name - an IDE plugin designed to rapidly and accurately identify inconsistencies in PL/SQL code during actual programming. We made \name publicly available online at  \url{https://github.com/JiangshanLiu/PLSQLIC3}.
    \item We conduct a user study with 16 volunteers of various PL/SQL experiences to evaluate the effectiveness of \name.
\end{itemize}


%% file: sections/1_Introduction_Figure_SQL_Injection.tex
\begin{figure}[t]
\lstset{
 tabsize=2,
 basicstyle=\tt\scriptsize,
 columns=fixed,       
 numbers=left,
 numbersep=4pt,
 numberstyle=\tiny\color{darkgray},
 frame=none,
 backgroundcolor=\color[RGB]{245,245,245},
 keywordstyle=\bfseries,
 commentstyle=\it\color[RGB]{0,96,96},
 showstringspaces=false,
 language=SQL,
 morekeywords={FUNCTION, DECLARE, IF, ELSIF, RETURN, FOR, LOOP, LANGUAGE, RETURNS, VOID, AS},
}
\begin{lstlisting}[]
CREATE FUNCTION reset(account_prefix CHAR) RETURNS VOID AS $$
BEGIN
  EXECUTE 'UPDATE users SET userpass = ''default'' WHERE 1 = '
    || account_prefix;
$$ LANGUAGE plpgsql;

SELECT * FROM reset('2');         -- Updates will not perform
SELECT * FROM reset('2 OR TRUE'); -- Updates performed by mistake,
                                  -- and reach SQL injection
\end{lstlisting}
\caption{SQL injection example caused by inconsistency. 
}
\label{fig:SQL-injection}
\end{figure} 

%% file: sections/2_PLSQL_IC3_Figure_Overview.tex
\begin{figure}[t]
    \centering
    \includegraphics[width=\linewidth]{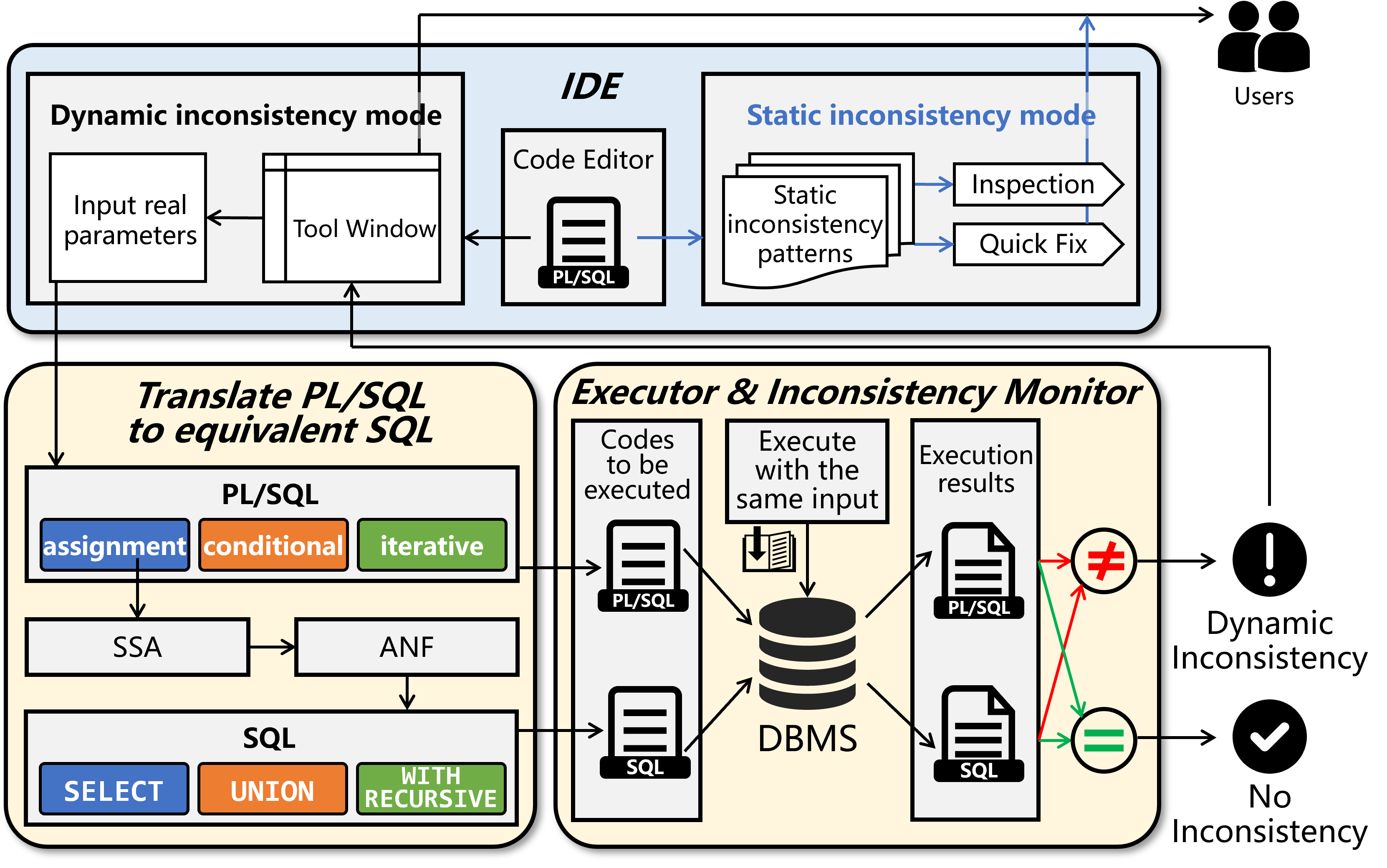}
    \caption{Overview of the method.}
    \label{fig:overview}
    \vspace{5pt}
\end{figure}

%% file: sections/2_PLSQL_IC3.tex

\section{Method}
\label{sec:method}

Inspired by the idea of metamorphic testing~\cite{chen2018metamorphic}, we propose to  translate a PL/SQL program to an equivalent SQL program, and then invoke the corresponding execution engines to execute the programs. Then we compare the execution results, which should be identical, to inspect inconsistencies. 
The intuition of our methods is two-fold. Firstly, since a large portion of PL/SQL programmers rely on their SQL experience for programming, we also use SQL as the reference language, which helps the PL/SQL programmers understand the real semantic of the PL/SQL programs they've written. Secondly, our method should be able to inspect the inconsistencies within the target DBMS itself, since unlike SQL, there is no standard specification for PL/SQL and thus different DBMS implementations that support PL/SQL language vary in both syntax and semantics of the PL/SQL they support. 
Figure \ref{fig:overview} shows the overview of our method, which consists of three components, i.e., translate PL/SQL to equivalent SQL, executor and inconsistency monitor and the IDE that interacts with the two components. 

\input{sections/3_IDE_Integration_Figure_IDEArch}

\input{sections/2_PLSQL_IC3_Table_Translation}
\input{sections/3_IDE_Integration_Figure_Demonstration}

\subsection{Translate PL/SQL to Equivalent SQL}\label{sec:mr}
We adopt and expand Hirn's work~\cite{hirn2021one, hirn2020pl, duta2019compiling}, which provides translation  rules from PL/SQL to SQL, and introduce rules that 
address 
PL/SQL syntax units such as {\ttfamily CASE-WHEN}, {\ttfamily ASSERT}, and cursor-style {\ttfamily FOR LOOP}. In essence, our translation rules translate PL/SQL into a literal SQL query\footnote{A literal  query's data source is a constant or variable, not a  table.}, where each variable in PL/SQL corresponds to a column in the SQL query. As the PL/SQL program progresses, each variable update is mapped to a new row in the literal query. The final state of the literal query (i.e., the last row) represents the final result of the PL/SQL execution. 
Our translation rules  concentrate on three types of  PL/SQL language features. 

\begin{itemize}
\item \textbf{Variable declaration and assignment} will be translated to a straightforward literal {\ttfamily SELECT} query.
\item \textbf{Conditional control} will be translated into two mutually exclusive condition queries integrated through {\ttfamily UNION ALL} in a consolidated query.
\item \textbf{Loop iteration} will be transformed into a recursive query by {\ttfamily WITH RECURSIVE} in SQL, a form of Common Table Expression (CTE) that iteratively queries until the termination condition is met.
\end{itemize}


As shown in Table \ref{tab:syntax-unit-MRs}, diverse PL/SQL syntactic units are categorised into three translation entities, whose corresponding equivalent translation rules are described through metamorphic relations. To make the translation process more intuitive, we also give examples of equivalent translations in our technical report~\cite{TechReport}.

\subsection{Executor \& Inconsistency Monitor}\label{sec:icmonitor}
After obtaining the equivalent SQL program from its equivalent PL/SQL program, we need to execute both of them to obtain the execution results. 
We execute PL/SQL and equivalent SQL with the same input parameters in the same environment. In particular, the SQL query invokes the SQL execution engine and the PL/SQL program invokes the PL/SQL execution engine of the same DBMS.  
If the execution results are different, we report the identified inconsistencies in the IDE views.



It has been shown that dynamically fixing semantic bugs is more challenging as the correct parameters are necessary to extract fix changes~\cite{liu2019avatar}. The same conclusion applies to our approach, since specific parameters are required to trigger inconsistencies.

To uncover more valuable inconsistencies, we employ the concept of fuzz testing~\cite{zhu2022fuzzing}, in which we collect real-world PL/SQL programs from public code repositories, open-source test case sets, and past academic work as seeds, mutate the seeds to obtain a large number of PL/SQL programs and execute them offline with our approach to uncover inconsistencies. 
The fuzzing process is executed automatically on the server, and when inconsistencies in the results of PL/SQL and equivalent SQL execution are monitored, a record is written to the log containing the name of the database and schema that was executed, the set of equivalent PL/SQL and SQL code pairs that were executed, the input parameters of the triggering code, and the results of the respective execution of the equivalent code pairs. We performed the fuzzing process for three months and analysed the automatically reported inconsistency log records. All the inconsistency records were distinguished into eight different inconsistencies, which belonged to three categories of issues.
Those inconsistencies are encoded as inconsistent patterns in the static mode of our tool, which does not trigger the execution engines and is more efficient. 



%% file: sections/3_IDE_Integration_Figure_IDEArch.tex
\begin{figure}[t]
\centerline{\includegraphics[width=\linewidth]{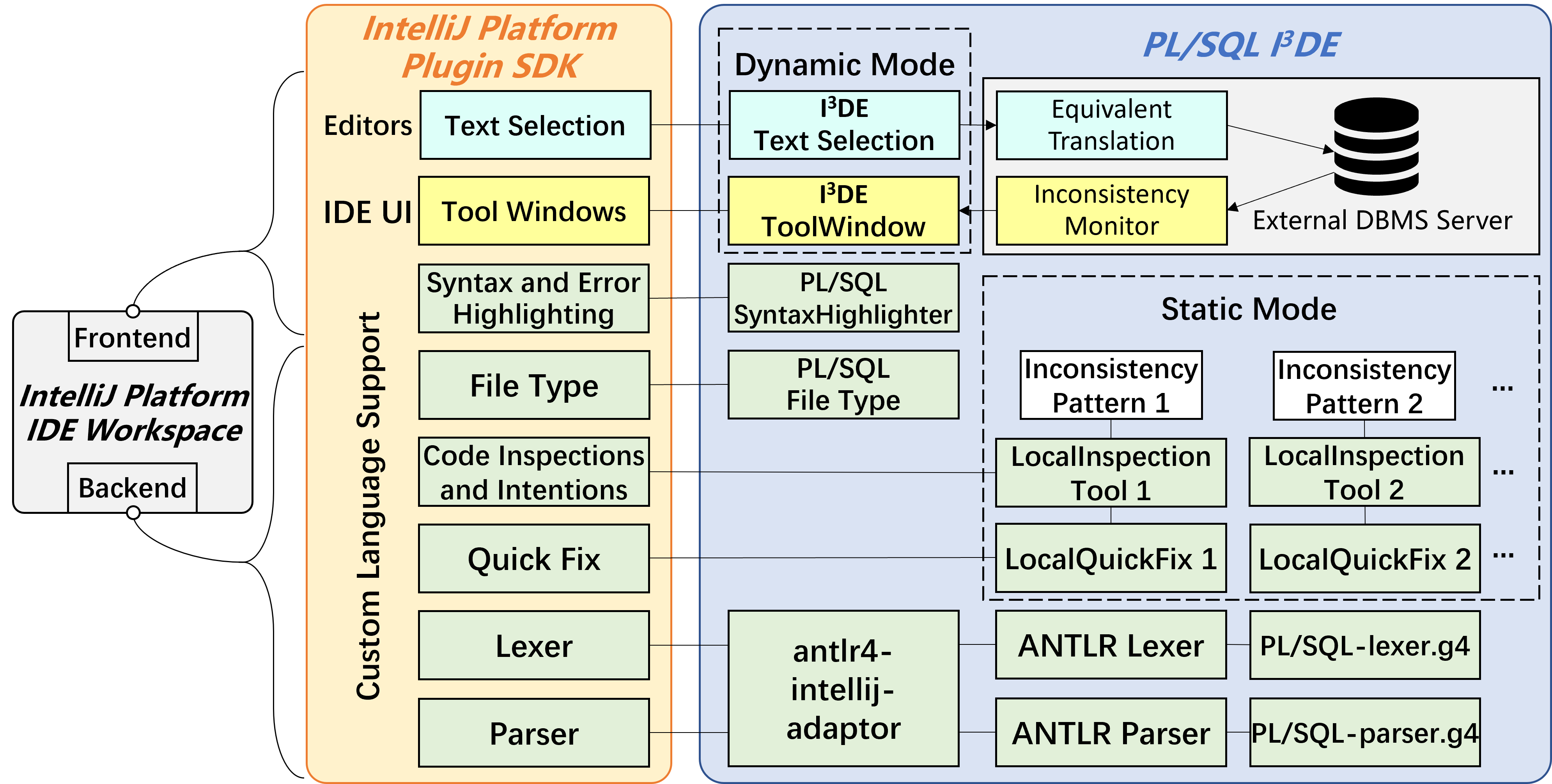}}
\vspace{2pt}
\caption{Architecture of IntelliJ Platform and \name.}
\label{fig:arch_ide}
\vspace{5pt}
\end{figure}

%% file: sections/2_PLSQL_IC3_Table_Translation.tex
\begin{table}[t]
\caption{Translation entities and their corresponding metamorphic relations.}
\label{tab:syntax-unit-MRs}
\renewcommand\arraystretch{1}
\begin{center}
\begin{tabular}{@{}c@{}c@{}c@{}}
\toprule
\textbf{\renewcommand\arraystretch{1} \begin{tabular}[c]{@{}c@{}}Translation\\ Entity\end{tabular}} & \textbf{\renewcommand\arraystretch{1} \begin{tabular}[c]{@{}c@{}}PL/SQL\\ Syntax Unit\end{tabular}} & \textbf{\renewcommand\arraystretch{1}  \begin{tabular}[c]{@{}c@{}}Metamorphic Relation for\\Equivalent Translation\end{tabular}} \\ \midrule
\multirow{3}{*}{\begin{tabular}[c]{@{}c@{}}variable\end{tabular}}                  &
declare                                                                & \multirow{3}{*}{\footnotesize $\dfrac{var \ type \ := \ value ;}{\mathtt{LATERAL}\ (\mathtt{SELECT}\ value::type)\ \mathtt{AS}\ \mathrm{tab}(var)}$} \\
&       assign                                                         &                                                                                                              \\
&       return                                                         &                                                                                                              \\ \midrule
\multirow{3}{*}{\begin{tabular}[c]{@{}c@{}}conditional\\ control\end{tabular}}     &
if-else                                                                & \multirow{3}{*}{\footnotesize $\dfrac{\mathtt{IF}\ cond\ \mathtt{THEN}\ stmt_1;\ \mathtt{ELSE}\ stmt_2;\ \mathtt{END\ IF};}{\setlength{\jot}{-0.2em} \begin{aligned} & \mathtt{SELECT}\ vars\ \mathtt{FROM}\ stmt_1\ \mathtt{WHERE}\ cond \\ & \quad \mathtt{UNION\ ALL} \\ &  \mathtt{SELECT}\ vars\ \mathtt{FROM}\ stmt_2\ \mathtt{WHERE}\ \mathrm{not}\ cond \end{aligned}}$}  \\
&       case-when                                                      &                                                                                                              \\
&       assert                                                         &                                                                                                              \\ \midrule
\multirow{7}{*}{\begin{tabular}[c]{@{}c@{}}loop\\ iteration\end{tabular}}          &
loop                                                                   & \multirow{7}{*}{\footnotesize $\dfrac{\setlength{\jot}{-0.12em} \begin{aligned} & \mathrm{\textit{LABEL}}_{init}: \\ & \quad stmt_{init}; \\ & \quad \mathtt{GOTO}\ \mathrm{\textit{LABEL}}_{body}; \\ & \mathrm{\textit{LABEL}}_{body}: \\ & \quad stmt_{body}; \\ & \quad \mathtt{IF}\ cond\ \mathtt{THEN}\ \mathtt{GOTO}\ \mathrm{\textit{LABEL}}_{body};\end{aligned}}{\setlength{\jot}{-0.2em} \begin{aligned} & \mathtt{WITH}\ \mathtt{RECURSIVE}\ \mathrm{run(\ )}\ \mathtt{AS}\ ( \\ & \quad \mathtt{SELECT}\ vars\ \mathtt{FROM}\ stmt_{init}\ \\ & \quad\quad \mathtt{UNION}\ \mathtt{ALL} \\ & \quad \mathtt{SELECT}\ vars\ \mathtt{FROM}\ stmt_{body}\ \mathtt{WHERE}\ cond)\end{aligned}}$}  \\
&       while                                                          &                                                                                                              \\
&       for                                                            &                                                                                                              \\
&       foreach                                                        &                                                                                                              \\
&       continue                                                       &                                                                                                              \\
&       exit                                                           &                                                                                                              \\
&       cursor\_for                                                    &                                                                                                              \\ \bottomrule
\end{tabular}
\end{center}
\end{table}

%% file: sections/3_IDE_Integration_Figure_Demonstration.tex

\begin{figure*}[t]
  \centering
  \begin{subfigure}[b]{0.49\textwidth}
    \includegraphics[width=\textwidth]{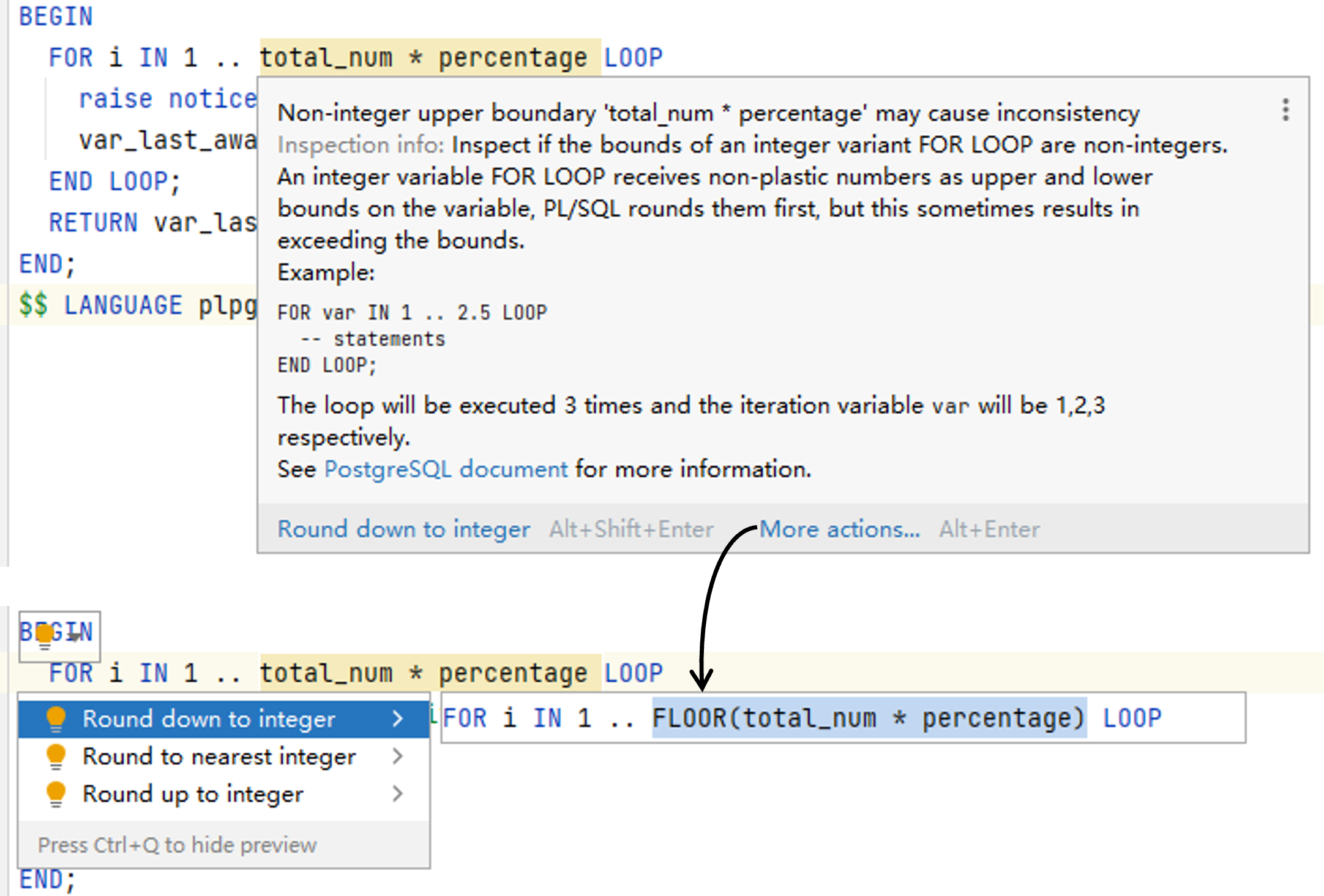}
    \caption{Static mode}
    \label{fig:demostatic}
  \end{subfigure}
  \hfill
  \begin{subfigure}[b]{0.49\textwidth}
    \includegraphics[width=\textwidth]{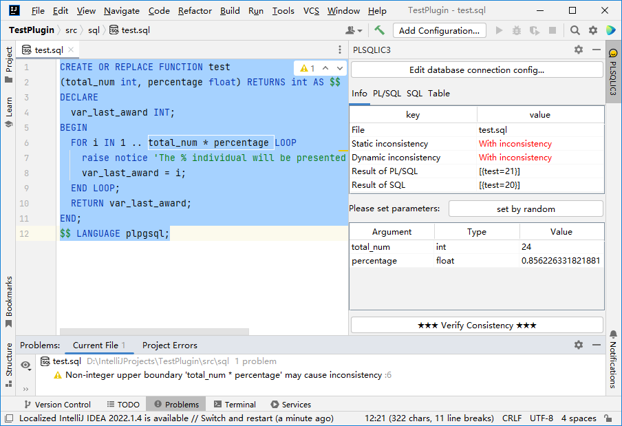}
    \caption{Dynamic mode}
    \label{fig:demodynamic}
  \end{subfigure}
\vspace{-2pt}
  \caption{Snapshots of \name static mode (a) and dynamic mode (b). 
  }
  \label{fig:demonstration}
\vspace{3pt}
\end{figure*}

%% file: sections/3_IDE_Integration.tex
\section{IDE Integration}\label{sec:IDE}

Figure \ref{fig:arch-ide} depicts the architecture of \name integration within IntelliJ, which consists of 3 parts, i.e., the IDE Workspace, the Plugin SDK provided by the IntelliJ Platform, and PL/SQL \name. 

\subsection{Overview of \name}
We develop \name as a plugin on the IntelliJ Platform, a highly extensible IDE framework known for its widespread compatibility with various IntelliJ-based IDEs\footnote{We developed and tested \name on IntelliJ IDEA, which is theoretically compatible with PyCharm, DataGrip and other IDEs.}~\cite{kurbatova2021intellij}. Within the IntelliJ Platform, the core IDE Workspace orchestrates frontend and backend tasks~\cite{IntelliJArch}, with the option for secondary development of these IDE tasks through interfaces provided by the Plugin SDK. \name redefines the PL/SQL file type and incorporates syntax and error highlighting, achieved through the PL/SQL lexer and parser implemented by us.



There are two modes in \name, i.e., static mode and dynamic mode. 
In static mode, \name utilizes the {\ttfamily LocalInspectionTool} from Plugin SDK to perform code inspections and intentions. Inspected inconsistencies are communicated to programmers through code highlighting and inspection descriptions. For fixable inconsistencies, {\ttfamily LocalQuickFix} is used to perform one-click code repair and preview.
For dynamic mode, \name registers a {\ttfamily ToolWindow} with the IntelliJ Platform to implement UI and interactions. The plugin captures the PL/SQL code to be inspected by listening to user-selected operations. After programmers set input parameter values in the {\ttfamily ToolWindow} and click submit, the PL/SQL code and input values are passed to the processing logic and triggers our method introduced in Section \ref{sec:method}. The execution results are compared for  inconsistencies, which are then returned to the {\ttfamily ToolWindow} for presentation to the programmer. 

\input{sections/4_Evaluation_Table_Participant}

\subsection{The Static Mode}

We run \name in dynamic mode for three months with the automated generated PL/SQL programs, as descried in Section \ref{sec:icmonitor}, and  collated the inconsistencies, based on which we predefined 8 corresponding static inconsistency patterns, which can be classified into 3 categories.  


\begin{itemize}
\item \textbf{Presumption. } Programmers assume that PL/SQL should have some kind of operation or processing, which is not the case.
For example, in line 4 of the code in Figure \ref{fig:SQL-injection}, when the input parameter type is \ttfamily{CHAR} with default length, the automatic type conversion is not executed, which is easy to be incorrectly predicted by the programmer.
\item \textbf{Overlook. }  Programmers ignore the operation or processing that PL/SQL engine implicitly does.
For example, in line 6 of the code in \ref{fig:demodynamic}, the \ttfamily{FOR} loop actually does rounding in the background when calculating the boundary value, and this implicit operation is easily ignored by the programmer.
\item \textbf{Equivocality. } The same keywords have different syntax and semantics in SQL and PL/SQL, and the programmer mistakenly uses one for the other.
For example, keywords such as \ttfamily{INTO}, \ttfamily{EXECUTE}, and \ttfamily{RETURNING} may have different semantics in SQL and PL/SQL.
\end{itemize}

For each pattern, carefully crafted programmer prompts and correction suggestions are designed to guide the programmer. 
Detailed information on these patterns can be found in our technical report~\cite{TechReport}.

As shown in Figure \ref{fig:demostatic}, when the PL/SQL code triggers predefined static inconsistency patterns, the programmer receives inconsistency warning messages without any additional actions. After understanding the inconsistency information, the programmer can decide whether to make changes based on the modification suggestions provided by \name. Thanks to IntelliJ's excellent code-fixing capabilities, this correction process is completed with a one-click operation.

\subsection{The Dynamic Mode}

\name{} also has a dynamic mode, in which the metamorphic testing-based method is triggered to inspect inconsistencies. 
As shown in Figure \ref{fig:demodynamic}, programmers need to select the PL/SQL code to be inspected in the editor and manually specify  the input parameters for it in the \name window. Subsequently, by clicking the inspect inconsistency button, the logic of the metamorphic testing based  inconsistency inspection method introduced in Section \ref{sec:method} will be triggered. The plugin backend connects to the  database service and executes the PL/SQL and its equivalent SQL. The existence of inconsistencies is then presented to the programmer based on the execution results from the database service. Programmers, upon discovering new inconsistencies, have the option to submit information triggering these inconsistencies to us through GitHub issues. 
The information will then be added in our patterns, which enables inspection of more types of inconsistencies in the static mode. 


%% file: sections/4_Evaluation_Table_Participant.tex


\begin{table}[t]
\caption{Proficiency with PL/SQL of each participant.}
\label{tab:participant}
\centering
\begin{tabular}{@{}ccccccccc@{}}
\toprule
\textbf{Group}         & \multicolumn{8}{c}{\textbf{Experimental group}} \\ \midrule
participant            & A    & C    & E   & G   & I   & K   & M   & O   \\
subjective rating      & 6    & 4    & 1   & 3   & 1   & 2   & 1   & 0   \\
objective score        & 8    & 6.5  & 8.5 & 6.5 & 7.5 & 4.5 & 4.5 & 2.5 \\
integrated proficiency & 7.0  & 5.3  & 4.8 & 4.8 & 4.3 & 3.3 & 2.8 & 1.3 \\ \midrule\midrule
\textbf{Group}         & \multicolumn{8}{c}{\textbf{Control group}}      \\ \midrule
participant            & B    & D    & F   & H   & J   & L   & N   & P   \\
subjective rating      & 3    & 6    & 2   & 3   & 1   & 1   & 1   & 0   \\
objective score        & 8.5  & 5.5  & 7.5 & 6.5 & 6   & 5.5 & 4.5 & 5   \\
integrated proficiency & 5.8  & 5.8  & 4.8 & 4.8 & 3.5 & 3.3 & 2.8 & 2.5 \\ \bottomrule
\end{tabular}
\vspace{5pt}
\end{table}

%% file: sections/4_Evaluation.tex
\section{Evaluation}
\label{sec:evaluation}


\input{sections/4_Evaluation_Table_Correctness}
\input{sections/4_Evaluation_Table_Time}

In order to examine whether \name can genuinely assist programmers in identifying and mitigating inconsistencies, we conducted a user study with 16 volunteers that have SQL and PL/SQL experiences. Table \ref{tab:participant} shows the proficiency in PL/SQL of 16 participants. We designed a questionnaire, which includes both subjective ratings on their proficiency in PL/SQL and objective questions accessing their proficiency in PL/SQL. 
%
The participants were then categorized into the experiment group  and the control group based on their proficiency levels, ensuring fair partition, i.e., diverse proficiency in each group and minimum difference between two groups. 
Three tasks representing different types of inconsistencies (detailed in our technical report~\cite{TechReport}) were designed to evaluate participants' ability to recognize and address inconsistencies in PL/SQL code. The experiment group  used the \name environment and received training on its usage, while the control group used the same IDE without \name but was  allowed to complete tasks using search engines, PL/SQL documentation, and generative language models. Responses and time duration for each task were recorded for analysis and evaluation.

\input{sections/7_Appendix_Figure_Overlook}

\noindent\textbf{RQ1: Is \name useful?} The correctness rate of task completion is presented in Table \ref{tab:correctness}. 
The control group achieved a correctness rate of 12.5\% for Tasks 1 and 3. Task 2, lacking a description of the inconsistency even in the documentation, failed all participants in the control group. 
With the assistance of \name, the experimental group achieved an average correctness rate of 95.83\%, representing a remarkable improvement of 87.5\% compared to the control group. Therefore, \name effectively aids programmers in recognizing and avoiding inconsistencies in PL/SQL.
Only 1 volunteer in the control group successfully identified the inconsistency in task 3 and provided a fix, with a  substantial time cost of 18 minutes 13 seconds.   
Furthermore, Task 2 involved an inconsistency with no relevant description even in the documentation, which makes it hard for manual inspection. 
We also tried using ChatGPT to complete the tasks and it fails on all three tasks. 
The results show  that \name does not require users with rich experience on PL/SQL documentation or pursue any external assistance in inspecting the inconsistencies.

\noindent\textbf{RQ2: Is \name efficient?} Task completion time is presented in Table~\ref{tab:time}. On average, the control group uses almost twice the time of the control group to finish all three tasks. 
As the difficulty of the tasks increase, i.e., from task 1 to task 3, the time saved with \name is more significant.
Moreover, with the increase difficulty of the tasks, the control group uses more time to finish the task, yet the experiment group shows a more stable time usage. 
The results indicate that \name is efficient in assisting programmers in identifying inconsistencies in programs with various level of difficulties.  



\noindent\textbf{RQ3: IS IDE a good solution for the problem?}
The feedback from participants in the experimental group indicates that the IDE, particularly on the IntelliJ Platform, played a crucial role in assisting the participants finishing the tasks. The inspection of potential code issues is automatically conducted in the background by the IntelliJ Platform engine~\cite{ketkar2022inferring, smirnov2022intellitc}, eliminating the need for additional user-initiated actions. This not only saves users considerable time but also effectively prevents potential catastrophic consequences resulting from users skipping operations that would trigger inspections. As plugin developers, we acknowledge that the IntelliJ Platform community provides comprehensive plugin development documentation and extensive development templates. This proves beneficial for researchers integrating their academic prototypes into the IDE for practical use. 
Therefore, we regard IDE as a suitable solution for the incosistency checking problem. 

%% file: sections/4_Evaluation_Table_Correctness.tex
\begin{table}[t]
\caption{
Correctness rate (\%) for each task.}
\label{tab:correctness}
\centering
\begin{tabular}{@{\hspace{0.3em}}lllll@{\hspace{0.3em}}}
\hline
                            & \textbf{Task 1} & \textbf{Task 2} & \textbf{Task 3} & \textbf{Average} \\ \hline
\textbf{Control group}      & 12.50\%         & 0.00\%          & 12.50\%         & 8.33\%           \\
\textbf{Experimental group} & 100.00\%        & 87.50\%         & 100.00\%        & 95.83\%          \\
\textbf{Improvement}        & 87.50\%         & 87.50\%         & 87.50\%         & 87.50\%          \\ \hline
\end{tabular}
\vspace{10pt}
\end{table}

%% file: sections/4_Evaluation_Table_Time.tex
\begin{table}[t]
\caption{The time (s) used to complete each task. 
}
\label{tab:time}
\centering
\begin{tabular}{@{\hspace{0.2em}}lllll@{\hspace{0.2em}}}
\hline
                                & \textbf{Task 1} & \textbf{Task 2} & \textbf{Task 3} & \textbf{Total} \\ \hline
\textbf{Control group}      & 131             & 197             & 433             & 762            \\
\textbf{Experimental group} & 115             & 127             & 141             & 383            \\
\textbf{Speedup} & 1.14         & 1.55         & 3.07         & 1.99        \\ \hline
\end{tabular}
\vspace{5pt}
\end{table}

%% file: sections/7_Appendix_Figure_Overlook.tex
\begin{figure}[t]
\lstset{
 tabsize=2,
 basicstyle=\tt\scriptsize,
 columns=fixed,       
 numbers=left,
 numbersep=4pt,
 numberstyle=\tiny\color{darkgray},
 frame=none,
 backgroundcolor=\color[RGB]{245,245,245},
 keywordstyle=\bfseries,
 commentstyle=\it\color[RGB]{0,96,96},
 showstringspaces=false,
 language=SQL,
 morekeywords={FUNCTION, DECLARE, IF, ELSIF, RETURN, FOR, LOOP, LANGUAGE, RETURNS, VOID, AS},
}
\begin{lstlisting}[]
-- Fill in the blank of the procedure.
-- The following procedure awards prizes to the top i individuals
-- by passing in the total_num and the percentage of the prize.
CREATE FUNCTION award(total_num int, percentage float)
  RETURNS int AS $$
DECLARE
  var_last_prize INT;  -- Incorrect: total_num * percentage
BEGIN                  -- correct: FLOOR(total_num * percentage)
  FOR i IN 1 .. _____ LOOP
    raise notice 'Prize for the person with ranking %', i;
    var_last_prize = i;
  END LOOP;
  RETURN var_last_prize;
END;
$$ LANGUAGE plpgsql;
\end{lstlisting}
\caption{User study task designed based on inconsistency caused by overlook.}
\label{fig:overlook_task}
\end{figure} 

%% file: sections/5_Related_Work.tex
\section{Related Work}

Little research focuses on programming language inconsistencies, mainly because unlike PL/SQL, most languages have unified specifications. Nuseibeh et al.~\cite{nuseibeh2001making} emphasize the inevitability of introducing inconsistencies in software development but stress their potential as triggers for constructive action. Consensus in the software ecosystem field highlights the importance of interoperability between projects within a large system, advocating the avoidance of inconsistencies~\cite{ma2020impact, manikas2016revisiting, manikas2013reviewing}.

Works addressing error prevention in IDEs are also related to our approach. Li et al.~\cite{li2019evaluation} evaluated five open-source IDE plugins for detecting security vulnerabilities, revealing high false positives and usability limitations. Amankwah et al.~\cite{amankwah2023bug} integrated eight automated static analysis tools into the Juliet Test Suite, demonstrating their effectiveness in detecting security bugs in Java source code.



%% file: sections/6_Conclusion.tex
\section{Conclusion}

In this paper, we introduce \name — an IDE plugin to inspect inconsistencies in PL/SQL code. Leveraging the concepts of fuzz testing and metamorphic testing, \name provides support for identifying and avoiding inconsistencies in two modes, i.e., static mode and dynamic mode. Static inconsistencies abstracted into predefined patterns are classified into three categories, with code inspection and quick fix in the IDE taking effect. Additionally, dynamic inconsistencies offer a runtime analysis with UI. \name is developed as a widely usable plugin for the IntelliJ Platform. We conducted a comparative user study involving 16 participants, and the findings indicate that \name is consistently effective and efficient in helping programmers identify and avoid inconsistencies across different programming difficulties.


%% file: main.bbl
\begin{thebibliography}{22}


\ifx \showCODEN    \undefined \def \showCODEN     #1{\unskip}     \fi
\ifx \showDOI      \undefined \def \showDOI       #1{#1}\fi
\ifx \showISBNx    \undefined \def \showISBNx     #1{\unskip}     \fi
\ifx \showISBNxiii \undefined \def \showISBNxiii  #1{\unskip}     \fi
\ifx \showISSN     \undefined \def \showISSN      #1{\unskip}     \fi
\ifx \showLCCN     \undefined \def \showLCCN      #1{\unskip}     \fi
\ifx \shownote     \undefined \def \shownote      #1{#1}          \fi
\ifx \showarticletitle \undefined \def \showarticletitle #1{#1}   \fi
\ifx \showURL      \undefined \def \showURL       {\relax}        \fi
\providecommand\bibfield[2]{#2}
\providecommand\bibinfo[2]{#2}
\providecommand\natexlab[1]{#1}
\providecommand\showeprint[2][]{arXiv:#2}

\bibitem[Amankwah et~al\mbox{.}(2023)]%
        {amankwah2023bug}
\bibfield{author}{\bibinfo{person}{Richard Amankwah}, \bibinfo{person}{Jinfu Chen}, \bibinfo{person}{Heping Song}, {and} \bibinfo{person}{Patrick~Kwaku Kudjo}.} \bibinfo{year}{2023}\natexlab{}.
\newblock \showarticletitle{Bug detection in Java code: An extensive evaluation of static analysis tools using Juliet Test Suites}.
\newblock \bibinfo{journal}{\emph{Software: Practice and Experience}} \bibinfo{volume}{53}, \bibinfo{number}{5} (\bibinfo{year}{2023}), \bibinfo{pages}{1125--1143}.
\newblock


\bibitem[Chen et~al\mbox{.}(2018)]%
        {chen2018metamorphic}
\bibfield{author}{\bibinfo{person}{Tsong~Yueh Chen}, \bibinfo{person}{Fei-Ching Kuo}, \bibinfo{person}{Huai Liu}, \bibinfo{person}{Pak-Lok Poon}, \bibinfo{person}{Dave Towey}, \bibinfo{person}{TH Tse}, {and} \bibinfo{person}{Zhi~Quan Zhou}.} \bibinfo{year}{2018}\natexlab{}.
\newblock \showarticletitle{Metamorphic testing: A review of challenges and opportunities}.
\newblock \bibinfo{journal}{\emph{ACM Computing Surveys (CSUR)}} \bibinfo{volume}{51}, \bibinfo{number}{1} (\bibinfo{year}{2018}), \bibinfo{pages}{1--27}.
\newblock


\bibitem[Duta et~al\mbox{.}(2019)]%
        {duta2019compiling}
\bibfield{author}{\bibinfo{person}{Christian Duta}, \bibinfo{person}{Denis Hirn}, {and} \bibinfo{person}{Torsten Grust}.} \bibinfo{year}{2019}\natexlab{}.
\newblock \showarticletitle{Compiling pl/SQL away}.
\newblock \bibinfo{journal}{\emph{arXiv preprint arXiv:1909.03291}} (\bibinfo{year}{2019}).
\newblock


\bibitem[Group(2023a)]%
        {PostgreSQLCHAR}
\bibfield{author}{\bibinfo{person}{The PostgreSQL Global~Development Group}.} \bibinfo{year}{2023}\natexlab{a}.
\newblock \bibinfo{title}{Character Types}.
\newblock \bibinfo{howpublished}{\url{https://www.postgresql.org/docs/current/datatype-character.html##DATATYPE-CHARACTER}}.
\newblock
\newblock
\shownote{accessed: November 2023}.


\bibitem[Group(2023b)]%
        {PostgreSQLOverview}
\bibfield{author}{\bibinfo{person}{The PostgreSQL Global~Development Group}.} \bibinfo{year}{2023}\natexlab{b}.
\newblock \bibinfo{title}{PL/pgSQL Overview}.
\newblock \bibinfo{howpublished}{\url{https://www.postgresql.org/docs/current/plpgsql-overview.html##PLPGSQL-OVERVIEW}}.
\newblock
\newblock
\shownote{accessed: November 2023}.


\bibitem[Group(2023c)]%
        {PostgreSQLPortingFromOracle}
\bibfield{author}{\bibinfo{person}{The PostgreSQL Global~Development Group}.} \bibinfo{year}{2023}\natexlab{c}.
\newblock \bibinfo{title}{Porting from Oracle PL/SQL}.
\newblock \bibinfo{howpublished}{\url{https://www.postgresql.org/docs/current/plpgsql-porting.html##PLPGSQL-PORTING}}.
\newblock
\newblock
\shownote{accessed: November 2023}.


\bibitem[Hirn and Grust(2020)]%
        {hirn2020pl}
\bibfield{author}{\bibinfo{person}{Denis Hirn} {and} \bibinfo{person}{Torsten Grust}.} \bibinfo{year}{2020}\natexlab{}.
\newblock \showarticletitle{PL/SQL Without the PL}. In \bibinfo{booktitle}{\emph{Proceedings of the 2020 ACM SIGMOD International Conference on Management of Data}}. \bibinfo{pages}{2677--2680}.
\newblock


\bibitem[Hirn and Grust(2021)]%
        {hirn2021one}
\bibfield{author}{\bibinfo{person}{Denis Hirn} {and} \bibinfo{person}{Torsten Grust}.} \bibinfo{year}{2021}\natexlab{}.
\newblock \showarticletitle{One with recursive is worth many GOTOs}. In \bibinfo{booktitle}{\emph{Proceedings of the 2021 International Conference on Management of Data}}. \bibinfo{pages}{723--735}.
\newblock


\bibitem[JetBrains(2023)]%
        {JetBrainsTools}
\bibfield{author}{\bibinfo{person}{JetBrains}.} \bibinfo{year}{2023}\natexlab{}.
\newblock \bibinfo{title}{JetBrains Tools for Data Science \& Big Data}.
\newblock \bibinfo{howpublished}{\url{https://www.jetbrains.com/data-tools/}}.
\newblock
\newblock
\shownote{accessed: November 2023}.


\bibitem[Ketkar et~al\mbox{.}(2022)]%
        {ketkar2022inferring}
\bibfield{author}{\bibinfo{person}{Ameya Ketkar}, \bibinfo{person}{Oleg Smirnov}, \bibinfo{person}{Nikolaos Tsantalis}, \bibinfo{person}{Danny Dig}, {and} \bibinfo{person}{Timofey Bryksin}.} \bibinfo{year}{2022}\natexlab{}.
\newblock \showarticletitle{Inferring and applying type changes}. In \bibinfo{booktitle}{\emph{Proceedings of the 44th International Conference on Software Engineering}}. \bibinfo{pages}{1206--1218}.
\newblock


\bibitem[Kurbatova et~al\mbox{.}(2021)]%
        {kurbatova2021intellij}
\bibfield{author}{\bibinfo{person}{Zarina Kurbatova}, \bibinfo{person}{Yaroslav Golubev}, \bibinfo{person}{Vladimir Kovalenko}, {and} \bibinfo{person}{Timofey Bryksin}.} \bibinfo{year}{2021}\natexlab{}.
\newblock \showarticletitle{The intellij platform: a framework for building plugins and mining software data}. In \bibinfo{booktitle}{\emph{2021 36th IEEE/ACM International Conference on Automated Software Engineering Workshops (ASEW)}}. IEEE, \bibinfo{pages}{14--17}.
\newblock


\bibitem[Li et~al\mbox{.}(2019)]%
        {li2019evaluation}
\bibfield{author}{\bibinfo{person}{Jingyue Li}, \bibinfo{person}{Sindre Beba}, {and} \bibinfo{person}{Magnus~Melseth Karlsen}.} \bibinfo{year}{2019}\natexlab{}.
\newblock \showarticletitle{Evaluation of open-source IDE plugins for detecting security vulnerabilities}. In \bibinfo{booktitle}{\emph{Proceedings of the 23rd International Conference on Evaluation and Assessment in Software Engineering}}. \bibinfo{pages}{200--209}.
\newblock


\bibitem[Liu(2023)]%
        {TechReport}
\bibfield{author}{\bibinfo{person}{Jiangshan Liu}.} \bibinfo{year}{2023}\natexlab{}.
\newblock \bibinfo{title}{Detecting inconsistencies in PL/SQL code through software testing methods}.
\newblock \bibinfo{howpublished}{\url{https://jiangshanliu.github.io/PLSQLIC3/web}}.
\newblock
\newblock
\shownote{accessed: November 2023}.


\bibitem[Liu et~al\mbox{.}(2019)]%
        {liu2019avatar}
\bibfield{author}{\bibinfo{person}{Kui Liu}, \bibinfo{person}{Anil Koyuncu}, \bibinfo{person}{Dongsun Kim}, {and} \bibinfo{person}{Tegawend{\'e}~F Bissyand{\'e}}.} \bibinfo{year}{2019}\natexlab{}.
\newblock \showarticletitle{Avatar: Fixing semantic bugs with fix patterns of static analysis violations}. In \bibinfo{booktitle}{\emph{2019 IEEE 26th International Conference on Software Analysis, Evolution and Reengineering (SANER)}}. IEEE, \bibinfo{pages}{1--12}.
\newblock


\bibitem[Ma et~al\mbox{.}(2020)]%
        {ma2020impact}
\bibfield{author}{\bibinfo{person}{Wanwangying Ma}, \bibinfo{person}{Lin Chen}, \bibinfo{person}{Xiangyu Zhang}, \bibinfo{person}{Yang Feng}, \bibinfo{person}{Zhaogui Xu}, \bibinfo{person}{Zhifei Chen}, \bibinfo{person}{Yuming Zhou}, {and} \bibinfo{person}{Baowen Xu}.} \bibinfo{year}{2020}\natexlab{}.
\newblock \showarticletitle{Impact analysis of cross-project bugs on software ecosystems}. In \bibinfo{booktitle}{\emph{Proceedings of the ACM/IEEE 42nd International Conference on Software Engineering}}. \bibinfo{pages}{100--111}.
\newblock


\bibitem[Manikas(2016)]%
        {manikas2016revisiting}
\bibfield{author}{\bibinfo{person}{Konstantinos Manikas}.} \bibinfo{year}{2016}\natexlab{}.
\newblock \showarticletitle{Revisiting software ecosystems research: A longitudinal literature study}.
\newblock \bibinfo{journal}{\emph{Journal of Systems and Software}}  \bibinfo{volume}{117} (\bibinfo{year}{2016}), \bibinfo{pages}{84--103}.
\newblock


\bibitem[Manikas and Hansen(2013)]%
        {manikas2013reviewing}
\bibfield{author}{\bibinfo{person}{Konstantinos Manikas} {and} \bibinfo{person}{Klaus~Marius Hansen}.} \bibinfo{year}{2013}\natexlab{}.
\newblock \showarticletitle{Reviewing the health of software ecosystems--a conceptual framework proposal}. In \bibinfo{booktitle}{\emph{Proceedings of the 5th international workshop on software ecosystems (IWSECO)}}. Citeseer, \bibinfo{pages}{33--44}.
\newblock


\bibitem[Nuseibeh et~al\mbox{.}(2001)]%
        {nuseibeh2001making}
\bibfield{author}{\bibinfo{person}{Bashar Nuseibeh}, \bibinfo{person}{Steve Easterbrook}, {and} \bibinfo{person}{Alessandra Russo}.} \bibinfo{year}{2001}\natexlab{}.
\newblock \showarticletitle{Making inconsistency respectable in software development}.
\newblock \bibinfo{journal}{\emph{Journal of systems and software}} \bibinfo{volume}{58}, \bibinfo{number}{2} (\bibinfo{year}{2001}), \bibinfo{pages}{171--180}.
\newblock


\bibitem[openGauss(2023)]%
        {openGaussOverview}
\bibfield{author}{\bibinfo{person}{openGauss}.} \bibinfo{year}{2023}\natexlab{}.
\newblock \bibinfo{title}{PL/pgSQL Functions}.
\newblock \bibinfo{howpublished}{\url{https://docs-opengauss.osinfra.cn/en/docs/latest/docs/SQLReference/pl-pgsql-functions.html}}.
\newblock
\newblock
\shownote{accessed: November 2023}.


\bibitem[Smirnov et~al\mbox{.}(2022)]%
        {smirnov2022intellitc}
\bibfield{author}{\bibinfo{person}{Oleg Smirnov}, \bibinfo{person}{Ameya Ketkar}, \bibinfo{person}{Timofey Bryksin}, \bibinfo{person}{Nikolaos Tsantalis}, {and} \bibinfo{person}{Danny Dig}.} \bibinfo{year}{2022}\natexlab{}.
\newblock \showarticletitle{IntelliTC: automating type changes in IntelliJ IDEA}. In \bibinfo{booktitle}{\emph{Proceedings of the ACM/IEEE 44th International Conference on Software Engineering: Companion Proceedings}}. \bibinfo{pages}{115--119}.
\newblock


\bibitem[s.r.o.(2023)]%
        {IntelliJArch}
\bibfield{author}{\bibinfo{person}{JetBrains s.r.o.}} \bibinfo{year}{2023}\natexlab{}.
\newblock \bibinfo{title}{Architecture overview}.
\newblock \bibinfo{howpublished}{\url{https://www.postgresql.org/docs/current/datatype-character.html##DATATYPE-CHARACTER}}.
\newblock
\newblock
\shownote{accessed: November 2023}.


\bibitem[Zhu et~al\mbox{.}(2022)]%
        {zhu2022fuzzing}
\bibfield{author}{\bibinfo{person}{Xiaogang Zhu}, \bibinfo{person}{Sheng Wen}, \bibinfo{person}{Seyit Camtepe}, {and} \bibinfo{person}{Yang Xiang}.} \bibinfo{year}{2022}\natexlab{}.
\newblock \showarticletitle{Fuzzing: a survey for roadmap}.
\newblock \bibinfo{journal}{\emph{ACM Computing Surveys (CSUR)}} \bibinfo{volume}{54}, \bibinfo{number}{11s} (\bibinfo{year}{2022}), \bibinfo{pages}{1--36}.
\newblock


\end{thebibliography}
